\documentclass[11pt,fleqn]{article}

\usepackage{amsfonts,amssymb,latexsym,cite}
\usepackage{amsmath, amssymb}
\usepackage{epstopdf}
\usepackage{eurosym}

\usepackage{amsfonts,amssymb,cite}
\usepackage{graphicx}

\def\noi{\noindent}

\newcommand{\Title}[1]{\noi {{\Large\bf #1}}\\[1ex]}

\def\Aunames#1{\noi{\bf #1}}
\def\auth#1{${}^{#1}$}
\def\Addresses#1{\medskip\noi \protect
	\begin{description}\itemsep -3pt {\it #1} \end{description}}
\def\addr#1#2{\item[${}^{#1}$]{\it #2}}

\newcommand{\Abstract}[1]{\vskip 2mm \begin{center}
        \parbox{16.4cm}{\small\noi #1} \end{center}\medskip}

\def\email#1#2{\footnotetext[#1]{e-mail: #2}\addtocounter{footnote}{1}}

\def\nqq{\hspace*{-2em}}
\def\nhq{\hspace*{-0.5em}}

\def\cm{\hspace*{1cm}}
\def\inch{\hspace*{1in}}

\def\Jl#1#2{#1 {\bf #2},\ }

\def\ApJ#1 {\Jl{Astroph. J.}{#1}}
\def\CQG#1 {\Jl{Class. Quantum Grav.}{#1}}
\def\DAN#1 {\Jl{Dokl. AN SSSR}{#1}}
\def\GC#1 {\Jl{Grav. Cosmol.}{#1}}
\def\GRG#1 {\Jl{Gen. Rel. Grav.}{#1}}
\def\JETF#1 {\Jl{Zh. Eksp. Teor. Fiz.}{#1}}
\def\JETP#1 {\Jl{Sov. Phys. JETP}{#1}}
\def\JHEP#1 {\Jl{JHEP}{#1}}
\def\JMP#1 {\Jl{J. Math. Phys.}{#1}}
\def\NPB#1 {\Jl{Nucl. Phys. B}{#1}}
\def\NP#1 {\Jl{Nucl. Phys.}{#1}}
\def\PLA#1 {\Jl{Phys. Lett. A}{#1}}
\def\PLB#1 {\Jl{Phys. Lett. B}{#1}}
\def\PRD#1 {\Jl{Phys. Rev. D}{#1}}
\def\PRL#1 {\Jl{Phys. Rev. Lett.}{#1}}

\def\al{&\nhq}
\def\lal{&&\nqq {}}
\def\eq{Eq.\,}
\def\eqs{Eqs.\,}
\def\beq{\begin{equation}}
\def\eeq{\end{equation}}
\def\bear{\begin{eqnarray}}
\def\bearr{\begin{eqnarray} \lal}
\def\ear{\end{eqnarray}}
\def\earn{\nonumber \end{eqnarray}}
\def\nn{\nonumber\\ {}}

\def\nnn{\nonumber\\ \lal }

\def\yy{\\[5pt] {}}
\def\yyy{\\[5pt] \lal }
\def\eql{\al =\al}

\def\dst{\displaystyle}
\def\tst{\textstyle}
\def\fracd#1#2{{\dst\frac{#1}{#2}}}
\def\fract#1#2{{\tst\frac{#1}{#2}}}
\def\Half{{\fracd{1}{2}}}
\def\half{{\fract{1}{2}}}

\def\e{{\,\rm e}}
\def\d{\partial}

\def\sign{\mathop{\rm sign}\nolimits}
\def\diag{\mathop{\rm diag}\nolimits}

\def\const{{\rm const}}

\def\ep{\epsilon}

\def\then{\ \Rightarrow\ }

\usepackage{color}

\def\mN{_\mu^\nu}

\def\eqn#1{\eq\eqref{#1}}
\def\rf{\eqref}

\def\GR{general relativity}
\def\EE{Einstein equation}

\def\KS{Kantowski-Sachs}

\def\ui{{\underline i}} 
\def\tT{{\widetilde T}}

\begin{document}
\twocolumn[

\Title{Duality between k-essence and Rastall gravity}

\Aunames{Kirill A. Bronnikov,\auth{a,b,c,1}
         	 J\'ulio C. Fabris,\auth{c,d,2}
	 	 Oliver F. Piattella,\auth{d}\\
		 Denis C. Rodrigues,\auth{d}
                   Edison C. Santos\auth{d}
	                }

\Addresses{
\addr a {VNIIMS, Ozyornaya ul. 46, Moscow 119361, Russia}
\addr b {Inst. of Gravitation and Cosmology, RUDN University,
         ul. Miklukho-Maklaya 6, Moscow 117198, Russia}
\addr c {National Research Nuclear University ``MEPhI'',
	 Kashirskoe sh. 31, Moscow 115409, Russia}
\addr d {Universidade Federal do Esp\'{\i}rito Santo,
 	 Vit\'oria, ES, CEP29075-910, Brazil}
	}

\Abstract
  {The k-essence theory with a power-law function of $(\d\phi)^2$ and Rastall's non-conservative 
  theory of gravity with a scalar field are shown to have the same solutions for the metric 
  under the assumption that both the metric and the scalar fields depend on a single coordinate. 
  This equivalence (called k-R duality) holds for static configurations with various symmetries 
  (spherical, plane, cylindrical, etc.) and all homogeneous cosmologies. In the presence of matter,
  Rastall's theory requires additional assumptions on how the stress-energy tensor non-conservation 
  is distributed between different contributions. Two versions of such non-conservation are considered
  in the case of isotropic spatially flat cosmological models with a perfect fluid: one (R1) in which 
  there is no coupling between the scalar field and the fluid, and another (R2) in which the fluid 
  separately obeys the usual conservation law. In version R1 it is shown that k-R duality holds not 
  only for the cosmological models themselves but also for their adiabatic perturbations. 
  In version R2, among other results, a particular model is singled out that reproduces the same
  cosmological expansion history as the standard $\Lambda$CDM model but predicts different 
  behaviors of small fluctuations in the k-essence and Rastall frameworks.
}
\medskip

] 
\email 1 {kb20@yandex.ru}
\email 2 {julio.fabris@cosmo-ufes.org}

\section{Introduction}

  The century-old general relativity (GR) theory still successfully passes 
  all local experimental tests. However, there are many reasons to consider 
  this theory not as an ultimate theory of gravity but only as a reasonable 
  approximation well working in a large but finite range of length and energy 
  scales. Among such reasons are the old problem of unifying gravity with other 
  physical interactions and the difficulties in attempts to quantize GR. 
  Other reasons for dealing with modifications of GR are the well-known problems 
  experienced by the theory itself: its prediction of space-time singularities
  in the most physically relevant solutions, actually showing situations where 
  the theory does not work any more, and its inability to explain the
  main observable features of the Universe without introducing so far invisible  
  forms of matter, dark matter (DM) and dark energy (DE), which 
  add up to as much as 95\,\% of the energy content of the Universe.
   
  The existing modifications and extensions of classical GR can be divided into 
  two large classes. The first one changes the geometric content of the theory
  and includes, in particular, $f(R)$ theories, multidimensional theories and 
  non-Riemannian geometries. The second class introduces new fundamental, 
  non-geometric fields and includes, in particular, scalar-tensor theories, 
  Horndeski theory \cite{horn} and vector-tensor theories. Of much interest are the 
  cases where  it is possible to establish connections between different representatives 
  of the same class or even different classes of theories (possibly the most well-known 
  example of such a connection is the equivalence of $f(R)$ theories with a 
  certain subclass of scalar-tensor theories, see, e.g., \cite{f(R),odin}). In the present paper 
  we discuss such an equivalence between large families of models of k-essence theories 
  and Rastall's non-conservative theories with a scalar field as a source of gravity. 

  The k-essence theories, introducing a non-stan\-dard form of the kinetic term 
  of a scalar field \cite{k1, k2}, evidently belong to the class of theories with
  non-standard fundamental fields coupled to gravity. They proved to be a way of 
  obtaining both early inflation and the modern accelerated expansion of the 
  Universe \cite{k2, k3, k4} driven by a scalar kinetic term instead of a 
  potential. Notably, a kind of k-essence structure also appears in string theories,
  for example, in the Dirac-Born-Infeld action, where the kinetic term of the 
  scalar field has a structure similar to that of the Maxwell-like term in Born-Infeld 
  electrodynamics \cite{dbi}.

  Rastall's theory \cite{rastall} is one more generalization of GR, which relaxes the
  conservation laws expressed by the zero divergence of the stress-energy tensor (SET) 
  $T\mN$ of matter. In this theory, the quantity $\nabla_\nu T\mN$ is linked to the 
  gradient of the Ricci scalar, and in this way Rastall's theory may be viewed as 
  a phenomenological implementation of some quantum effects in a curved background.
  Rastall's theory leads to results of interest in cosmology, e.g., the evolution of 
  small DM fluctuations is the same as in the $\Lambda$CDM model, but DE is able to 
  cluster. This might potentially provide an evolution of DM inhomogeneities in the non-linear 
  regime different from the standard CDM model \cite{cosmo}. The whole success of the 
  $\Lambda$CDM model is reproduced at the background and linear perturbation levels, 
  but new effects are expected in the non-linear regime, where the $\Lambda$CDM model
  faces some difficulties \cite{cusp, satt}. It has also been shown  
  \cite{oliver} that Rastall's theory with the canonical SET of a scalar field, in the context of  
  cosmological perturbations,  is only consistent if matter is present. An interesting 
  observation in this analysis is that scalar field coupling with gravity leads to equations 
  very similar to those in some classes of Galileon theories.

  A consideration of static, spherically symmetric solutions in k-essence theories with a power law
  kinetic function \cite{denis}
  and similar solutions in Rastall's theory in the presence of a free or self-interacting scalar field
  \cite{we-16} has shown that some exact solutions of these two theories describe quite the same  
  geometries, although the properties of the scalar fields are different. 
  Also, a no-go theorem concerning the possible emergence of Killing horizons, proved in the 
  k-essence framework \cite{denis}, has its counterpart in the Rastall-scalar field system
  \cite{we-16}. These similarities indicate a deeper relationship between the two theories,
  to be analyzed in this paper. We will show that in the absence of other matter than the 
  (possibly self-interacting) scalar fields, the two theories lead to completely coinciding 
  geometries (we will call this {\sl k-R duality}) under the assumptions that the relevant 
  quantities depend on a single spatial or temporal coordinate and that the k-essence theory 
  is specified by a power-law function; then, there emerges a simple 
  relation between the numerical parameters of the theories. If there are other forms of matter, 
  the situation is more involved and depends on how the non-conservation of the SET is 
  distributed between different matter contributions in Rastall's theory. We will discuss two 
  variants of such non-conservation in the case of isotropic spatially flat cosmological models
  and show that k-R duality generically takes place. 
   
  The paper is organized as follows. In the next section we discuss vacuum solutions of 
  k-essence and Rastall theories. In Section 3, isotropic cosmological models are analyzed 
  with a matter source in the form of a perfect fluid. Some considerations on the speed of 
  sound are presented in Section 4, while Section 5 is devoted to some special values of 
  the numerical parameters of both theories. Some concrete cosmological configurations 
  with dustlike matter are discussed in Section 6. Our conclusions are presented in Section 7.

\section{Scalar-vacuum space-times}

\subsection{k-essence}

  The k-essence theories can be defined as \GR\ with generalized forms 
  of scalar fields minimally coupled to gravity. In the absence of matter 
  nonminimally coupled to gravity, the most general Lagrangian is
\bearr 			\label{L-k}
	{\cal L} = \sqrt{-g}[R + F(X,\phi) + L_m],
\ear
  with
\bear                             \label{X}
	X = \eta\phi_\mu\phi^\mu,
\ear
  where $\phi_\mu = \d_\mu \phi$, $F(X,\phi)$ is an arbitrary function, and
  $\eta = \pm 1$ is used to make $X$ positive since otherwise in the cases
  like general power-law dependence $F$ will be ill-defined for $X < 0$;
  $L_m$ is the Lagrangian density of other kinds of matter having no direct 
  coupling to the curvature or the $\phi$ field.
  We are using the system of units where $c = 8\pi G =1$

 Variation of the Lagrangian (\ref{L-k}) with respect to the metric and
 the scalar field leads to the field equations
\bearr                                   		    \label{EE}
	G\mN \equiv R\mN - \half\delta\mN R = -T\mN [\phi] - T\mN[m],
\yyy                                                      \label{T-phi}
	T\mN [\phi] \equiv \eta F_X \phi_\mu \phi^\nu
		- \frac{1}{2} \delta\mN F ,
\yyy   				    		 \label{eq-phi}
       \eta\nabla_\alpha (F_X \phi^\alpha) - \Half\, F_\phi = 0,
\ear
  where $G\mN$ is the Einstein tensor, $F_X = \d F/\d X$, $F_\phi = \d F/\d\phi$,
  and $T\mN[m]$ is the SET of matter due to $L_m$.

  Now, let us make the following assumptions: \yy
{\bf (i)} the k-essence Lagrangian is
\beq                                              \label{L-k1}
          F (X, \phi) = F_0 X^n - 2V(\phi),
\eeq
  where $n = \const \ne 0$ and $V(\phi)$ is an arbitrary function (the potential).
\yy
{\bf (ii)} $\phi = \phi(u)$, where $u$ is one of the coordinates, which may be temporal or spatial.
\yy
{\bf (iii)} The metric has the form 
\beq                  \label{ds}
           ds^2 = \eta \e^{2\alpha(u)} du^2 + h_{ik} dx^i dx^k,  
\eeq 
  where $i, k$ are the numbers of coordinates other than $u$, and the determinant of $h_{ik}$
  has the factorized structure
\beq                                     \label{h_ik}
            \det (h_{ik}) = \e^{2\sigma(u)} h_1(x^i).
\eeq

  In this case, we have $X = \e^{-2\alpha(u)} \phi_u^2$ (the index $u$ means $d/du$), 
  and the SET of the $\phi$ field has the following nonzero components:
\bearr            \label{uu-k}
              T^u_u [\phi] =  (n - \half) F_0 X^n + V,
\yyy               \label{ii-k}
              T^\ui_i [\phi] = - \half F_0 X^n + V          
\ear
  (there is no summing over an underlined index). The scalar field equation has the form
\bearr                     \label{eq-k1}
             \e^{-2n\alpha} \phi_u^{2n-2} \big[(2n-1)\phi_{uu} + \sigma_u \phi_u  
\nnn \cm
                               - (2n-1) \alpha_u\phi_u \big] = - \frac {1}{nF_0} \frac {d V}{d\phi}.
\ear

\subsection{Rastall's theory with a scalar field}

  Rastall's theory of gravity is characterized by the following equations \cite{rastall}:
\bearr                                                                        \label{EE-Ra}
	R\mN - \frac{\lambda}{2}\delta\mN R = - T\mN,
\yyy
	\nabla_\nu T\mN = \frac{\lambda - 1}{2}\d_\mu R,
\ear
  where $\lambda$ is a free parameter and $T\mN$ is the SET of matter. 
  At $\lambda = 1$, GR is recovered.

  These equations can be rewritten as
\bearr                     	\label{EE-R1}
	G\mN = - \biggr\{T\mN - \frac{b - 1}{2}\delta\mN T\biggl\} \equiv - \tT\mN,
\yyy\nhq                              \label{cons-R} 
          \nabla_\nu T\mN =  \frac{b {-} 1}{2}\d_\mu T,\quad 
                  b: = \frac{3\lambda - 2}{2\lambda - 1}, \quad T = T^\alpha_\alpha.
\ear
  In this parametrization, GR is recovered if $b = 1$.

  Let us consider matter in the form of a minimally coupled scalar field $\psi$, so that  
\beq                           \label{SET-psi}
            T\mN[\psi] = \ep (\psi_\mu \psi^\nu - \half \delta\mN \psi_\alpha \psi^\alpha)
                                 + \delta\mN W(\psi),  
\eeq
  where $\ep = \pm 1$, indicating an ordinary ($+1$) or phantom ($-1$) nature 
  of the $\psi$ field, $\psi_\mu \equiv \d_\mu\psi$, and $W(\psi)$ is a potential. 
  The scalar field equation follows from \rf{cons-R} and has the form
\beq                                          \label{e-psi-R}
            \Box\psi + (b - 1)\frac{\psi^\mu \psi^\nu \nabla_\mu \psi_\nu}
                     {\psi^\alpha \psi_\alpha} = - \ep (3 - 2b) \frac {dW}{d\psi}.
\eeq
  
  Let us now, in full similarity with what was done for k-essence theory, assume that 
  $\psi = \psi(u)$ and the metric has the form \rf{ds}. Then the nonzero components 
  of the modified scalar field SET in the right-hand side of \eqn{EE-R1} are
\bearr                         \label{uu-R}
             \tT^u_u [\psi] = \half \ep b \eta \e^{-2\alpha} \psi_u^2 + (3-2b) W(\psi),
 \yyy                          \label{ii-R}
	    \tT^{\ui}_i [\psi] = \half \ep (b-2) \eta \e^{-2\alpha} \psi_u^2 + (3-2b) W(\psi),
\ear  
  while the scalar field equation \rf{e-psi-R} takes the form
\beq                                     \label{epsi-R2}
            \e^{-2\alpha}\big[b \psi_{uu} + \psi_u (\sigma_u - b\alpha_u)\big]
                                  = - \ep \eta (3-2b) W_\psi,
\eeq
  where $W_\psi \equiv dW/d\psi$ and, as before, $\eta = \sign g_{uu}$.

\subsection{Comparison}

  We assume that in the k-essence system there is no other matter than the scalar 
  field $\phi$ and in the Rastall system there is no other matter than the scalar $\psi$. 
  Let us find out under which conditions the right-hand sides of the \EE s \rf{uu-k}
  and \rf{ii-k} coincide with those of the effective \EE s of Rastall's theory,
  \rf{uu-R} and \rf{ii-R}. This will guarantee that the solutions for the metric are 
  also the same. 

  To begin with, we identify the potentials:
\beq         \label {VW}
                 V(\phi) = (3-2b) W(\psi).
\eeq 
  Next, we equate the ratios of the kinetic parts of $T\mN[\phi]$ and $\tT\mN[\psi]$,
  to obtain 
\beq             \label{nb}
      \frac{2n-1}{-1} = \frac b {b-2} \ \then \  (2-b) n = 1.  
\eeq
  Then, equating the kinetic parts themselves, we find that
\bearr              \label{phips}
           \psi^2_u = \ep\eta n F_0 \phi_u^{2n}\e^{2(1-n)\alpha}.
\ear 

  Under the three conditions \rf{VW}--\rf{phips}, the metric field equations of the 
  two theories completely coincide, therefore their sets of solutions are also identical.
  Substituting \rf{phips} to \rf{epsi-R2}, one can easily verify that under these conditions 
  the scalar field equations  \rf{eq-k1} and \rf{epsi-R2} are also equivalent.

  This general result covers many static symmetries (spherical, plane, cylindrical, etc.),
  homogeneous cosmologies (FRW, all Bianchi types, \KS) and even inhomogeneous 
  ones if their metrics are of the form \rf{ds}, \rf{h_ik}.

  Here and in most of the paper we consider the generic values of the parameters $n$ and 
  $b$ and exclude from consideration their special values that require a separate analysis, 
  such as, for example, $b=0$, $b=3/2$ and $n = 1/2$. Some remarks on these special cases 
  will be made in Section 5.   

\section{Cosmology with matter}
  
  When, besides the scalar field, matter is present, it is better, for evident technical reasons, 
  to restrict ourselves from the beginning to a certain type of metrics. We will consider 
  cosmological FLRW spatially flat metrics
\beq                     \label{ds1}
	  ds^2 = dt^2 - a(t)^2[dx^2 + dy^2 + dz^2],
\eeq
  so that in \rf{ds} and \rf{h_ik} we have $\eta=1$, $\e^\alpha =1$, and $\e^\sigma = a(t)^3$.
  Matter will be taken in the form of a perfect fluid, so that 
\beq               \label{Tm}
	  T\mN [m] = \diag (\rho, -p, -p, -p), 
\eeq  
  where $\rho$ is the density and $p$ is the pressure.

\subsection{k-essence cosmology}
  
 In the FLRW metric \rf{ds1} and with $\phi=\phi(t)$, the field equations \rf{EE}--\rf{eq-phi} 
 with matter (where we denote $\rho = \rho_k,\ p=p_k$) take the form
\bearr          \label{tt-k1}
  	3H^2 = \half(2n - 1) F_0\dot\phi^{2n} + V(\phi) + \rho_k,
\yyy            \label{ii-k1}
          2\dot H + 3H^2 = - \half F_0 \dot\phi^{2n} + V(\phi) - p_k,
\yyy            \label{phi-k1} 
         \dot\phi^{2n-2} [(2n-1)\ddot\phi + 3H\dot\phi] = - \frac{1}{nF_0}V_\phi,
\ear
  where $H =\dot a/a$ is the Hubble parameter and $V_\phi \equiv dV/d\phi$. The SET 
  of matter satisfies the conservation law $\nabla_\nu T\mN[m] =0$, whence
\beq               \label{cons}
	\dot\rho_k + 3H (\rho_k + p_k) =0.
\eeq 
  
\subsection{Rastall cosmology with a scalar field and matter}
\def\trho{{\widetilde \rho}}
\def\tp{{\widetilde p}}

  The Rastall equations have the form \rf{EE-R1} and \rf{cons-R}, where now  
  $T\mN$ is the total energy-momentum tensor,
\bear
              T\mN = T\mN [\psi] + T\mN [m],
\ear
  with $T\mN[\psi]$ given by \rf{SET-psi} and $T\mN[m]$ by \rf{Tm}. 
  In \eqs \rf{EE-R1}, the modified energy-momentum tensor $\tT\mN$ is then a sum
  of $\tT\mN[\psi]$ given by \rf{uu-R} and \rf{ii-R} and $\tT\mN[m]$ with the components
\bearr        \label {redef}
	  \tT^t_t [m] = \Half [(3-b) \rho + 3 (b-1) p] \equiv  \trho,
\nnn
	  \tT^\ui_i [m] = \Half [(1-b)\rho + (3b-5)p] \equiv - \tp 
\ear 
  (we preserve the notation $\rho$ and $p$ without indices for matter in Rastall gravity). 
  Hence the Rastall equations read
\bearr               \label{tt-Ra}
           3 H^2 = \Half \ep b \dot\psi{}^2 + (3-2b) W + \trho,
\yyy                  \label{ii-Ra}
	 2\dot H + 3H^2 = \Half \ep (b-2)  \dot\psi{}^2 + (3-2b) W - \tp,
\ear
  while the equation for $\psi$ depends of further assumptions on how the nonconservation 
  of the full SET according to \rf{cons-R} is distributed between $\psi$ 
  and matter. One can notice that 
\beq                 \label{NEC-R}
	\trho + \tp = \rho + p,
\eeq
  and 
\beq                              \label{trace}
        \trho - \rho = p - \tp = \frac{1-b}{2} (\rho - 3p).   
\eeq

  Let us consider two (of an infinite number of) alternatives in incorporating matter 
  to Rastall's theory with a scalar field:  
\begin{description}
\item[R1:] 
  The SETs of $\psi$ and matter obey \rf{cons-R} each separately, so there is
  no mixing between the two sources of gravity;
\item[R2:]
   The SET of matter is conservative, so that 
\beq 
            \dot\rho + 3H (\rho+p) =0.               
\eeq
\end{description}

\subsection{Case R1: No mixing of scalar field and matter} 

  In this case we have
\bearr              \label{cons-i1}
          \nabla_\nu T\mN[\psi] =  \frac{b - 1}{2}\d_\mu T[\psi],
\yyy                   \label{cons-i2}
          \nabla_\nu T\mN[m] =  \frac{b - 1}{2}\d_\mu T[m],
\ear
  The first of these conditions leads to the scalar field equation \rf{epsi-R2}, which in  
  the present case reads
\beq               \label{epsi-i} 
	b \ddot\psi + 3H \dot\psi = -\ep (3-2b) W_\psi.
\eeq
  With \rf{NEC-R}, the condition \rf{cons-i2} has the form 
\beq               \label{cons-i}
	\dot\trho + 3H (\rho + p) =0.
\eeq 
  The full set of equations consists of \rf{tt-Ra}, \rf{ii-Ra}, \rf{epsi-i}, and \rf{cons-i}, 
  with the definitions \rf{redef}.
  
  From \rf{NEC-R} it follows that if matter satisfies the null energy condition (NEC),
  then the same is true for the ``effective'' density and pressure ($\trho$ and $\tp$) in 
  Rastall's theory. However, from \rf{redef} it can be verified that the positivity of 
  the energy density (or pressure) is not guaranteed in the k-essence case if it is 
  imposed in the Rastall theory, and vice versa.

  It is easy to see that the right-hand sides of \eqs \rf{tt-Ra} and \rf{ii-Ra} coincide with
  those of \rf{tt-k1} and \rf{ii-k1} if, in addition to the relationships \rf{VW}--\rf{phips}
  for scalar variables, we identify 
\beq                                      \label{id-m}
                \rho_k = \trho, \qquad p_k = \tp.
\eeq
  The correctness of this identification is confirmed by the identity of the conservation 
  laws \rf{cons} and \rf{cons-i}. Thus, as in the vacuum case, the parameters 
  $n$ and $b$ are related by \rf{nb}, that is, $n(2-b) =1$, and the scalar fields $\phi$ and 
  $\psi$ are related by \eqn{phips} which now reads
\beq                                              \label{psif}
               \dot\psi{}^2 = \ep n F_0 \dot\phi{}^{2n}.
\eeq
  
\subsection{Case R2: Conservative matter}
 
  We now have $\nabla_\nu T\mN [m] = 0$. This condition is particularly important 
  for the structure formation in the universe for the case of a pressureless fluid 
  since ordinary matter must agglomerate.

  In this case, for the scalar field SET we have 
\bear
         \nabla_\nu T\mN [\psi] =\Half (b - 1) (\d_\mu T[\psi] + \d_\mu T[m]),
\ear
  which leads to the scalar field equation
\bearr              \label{epsi-ii} 
	\dot\psi(b \ddot\psi + 3H \dot\psi) = -\ep (3-2b) W_\psi \dot\psi
\nnn \inch
	+ \Half\ep (b-1)(\dot\rho - 3\dot p). 
\ear
  The full set of equations consists of \rf{tt-Ra}, \rf{ii-Ra}, \rf{epsi-ii}, and \rf{cons}, 
  with the definitions \rf{redef}. Note that \eqn{epsi-ii} mixes the scalar field and the 
  matter fluid even though the fluid is conserved as in GR.

  This conservation makes us identify the matter SET components in the k-essence
  and Rastall theories: $\rho = \rho_k$, $p = p_k$. As a result, identification of the other 
  parts of the total SET is only partly the same as in the previous case. 

  Identifying, as before, the potentials according to \rf{VW} (that is, $V = (3-2b)W$)
  and comparing the Friedmann-like equations \rf{tt-Ra} and \rf{ii-Ra} with 
  their k-essence counterparts \rf{tt-k1} and \rf{ii-k1}, we obtain, as before,  
\beq                                  \label{psif-ii}
                   \ep\dot\psi^2 = nF_0\dot\phi^{2n}.
\eeq
  The correctness of this identification is verified by substituting \rf{psif-ii} into 
  the scalar field equation \rf{epsi-ii}: indeed, since we have now, due to \rf{trace}, 
\beq                                   \label{nb-ii}
                   \ep\dot\psi^2 \Big[b - \frac{2n-1}{n}\Big] = (b-1) (\rho-3p),  
\eeq
  this substitution leads precisely to the scalar field equation \rf{phi-k1} of the 
  k-essence theory.

  It is important that in the case of conservative matter, a comparison 
  between the two theories does not lead to a direct relationship like 
  \rf{nb} between their numerical parameters $n$ and $b$. Instead, we have 
  the equality \rf{nb-ii}, from which \rf{nb} is restored only in the special case 
  $\rho=3p$ (zero trace of the matter SET, radiation).

\subsection{Further consequences of matter conservation}

  The relation \rf{nb-ii} creates a connection between the temporal behavior of $\psi$ and the 
  matter content. Indeed, inserting \rf{nb-ii} to \rf{epsi-ii} with zero or constant potential, 
  we find
\beq
                  \dot F + \frac{6n}{2n - 1}HF = 0,
\eeq
  where $F = \dot\psi^2$. This leads to
\beq             \label{sol1}
                 \dot\psi^2 = \psi_0a^{-6n/(2n - 1)},
\eeq
where $\psi_0$ is an integration constant.

  From \eq \rf{nb-ii} it is clear that the matter density and pressure  must also evolve by a 
  power law as functions of the scale factor. Hence, only an equation of state (EoS) of the type 
  $p = w\rho$, with $w = \const$, is possible. In this case, 
\beq
          \rho = \rho_0 a^{-3(1 + w)},      \qquad \rho_0 = \const,     
\eeq
  implying the relation between $w$ and $n$
\beq                                \label{rela}
               w = \frac{1}{2n - 1}.
\eeq
  We see that a substitution of  $p = w\rho$ into the scalar field equation relates the EoS factor
  $w$ with the k-essence power $n$, while the Rastall constant $b$ remains arbitrary. 
  Moreover, \eqs \rf{tt-k1} and \rf{ii-k1} show that the pure k-essence scalar field $\phi$ behaves 
  as a perfect fluid with the same EoS factor \rf{rela} (see also \cite{carla}).

  In other words, assuming a zero or constant potential $V = (3-2b)W$ and conservative matter 
  in the Rastall framework, we obtain that {\sl the k-R duality is only possible if matter 
  is a perfect fluid with the linear EoS $p = w\rho$, coinciding with the effective EoS
  of the scalar field $\phi$. }

  With $p = w\rho$ \eqn{nb-ii} gives 
\bearr
      \ep \dot\psi^2 = k\rho,
\qquad
	k = \frac{n(b- 1)(1 - 3w)}{bn-2n+1}.
\ear
  Inserting this to the Friedmann-like equation \rf{tt-Ra}, we obtain in terms of $n$ or $w$
\bear
            3H^2 \eql V + \frac{\rho}{2n - 1}
	\biggr\{\frac{2nb(b - 1)(n - 2)}{bn -2n + 1}
\nnn \cm\cm
	           + 2(2b - 3) + 2n(3 - b)\biggr\}
\yy 
           \eql   V + \frac{\rho}2 \biggl\{\frac{b(b-1)(1+w)(1-3w)}{(b-2)(1+w)+2w}
\nnn \cm\cm
                 + 3-b + 3w(b-1)\biggr\}.
\ear
  The right-hand side must be positive. Therefore, given $n$ and $V$ (or alternatively $w$ 
  and $V$), we obtain a restriction on $b$. For example, if $V=0$ and $w = 0$ 
  (dust, $n \to \infty$),\footnote
	{This relation makes sense even if the Lagrangian formulation becomes ill-defined,
	 see \cite{carla}.} 
  we have either $b < 3/2$ or $b > 2$ (provided $\rho > 0$). For $w = 1$ (stiff matter, $n = 1$), 
  there is no restriction on $b$, and we obtain $H^2 = V = \const >0$, hence a de Sitter 
  expansion, $a(t) \propto\e^{Ht}$. In this case, stiff matter precisely cancels the contribution 
  from the scalar field $\psi$ or $\phi$. 

  If we introduce a variable potential or a more complex equation of state, the situation becomes 
  much more involved. It must be stressed that the EoS $p = w\rho$ with $w=\const$
  covers most of the interesting cases in cosmology. Moreover, we expect that the 
  perturbative behavior may be very different in the two theories even in this case.

  There emerge two more natural questions. First, we have obtained that in k-essence theory
  there are simultaneously a scalar field and a perfect fluid with the same EoS and hence the same 
  time evolution of their densities and pressures. Can we unify them by, for example, redefining
  the scalar field? A probable answer is ``no'' because these two kinds of matter are expected to
  behave quite differently at the perturbative level.

  Another question is: how is it possible to have a completely definite situation in k-essence 
  theory but an arbitrariness in the parameter $b$ in the dual solution of Rastall's theory? An 
  answer is that this arbitrariness is compensated by the corresponding non-conservative 
  behavior of the scalar field $\psi$.     

\section{Perturbations and the speed of sound}

  A power law k-essence model with $V = 0$ is equivalent in the cosmological 
  framework (such that $(\d_\mu \phi)^2 > 0$) to a perfect fluid with the equation 
  of state $p = w \rho$, where the constant $w$ is related to the power $n$:
\beq      \label{w}
	w = \frac{1}{2n - 1}.
\eeq
  In a fluid, adiabatic perturbations propagate as a sound with the speed $v_s$ such that
\beq             \label{v-fl}
	v_s^2 = dp/d\rho = w.
\eeq
  
  Scalar field perturbations for general Lagrangians of the form $F(X,\phi)$
  have been treated in detail in \cite{neven, k3}. It has been shown there that
  a k-essence theory implies 
\beq
             v_s^2 = \frac{F_X}{F_X + 2X F_{XX}},
\eeq
  and this expression is valid even if there is an arbitrary potential term $V(\phi)$.
  In particular, for the theory \rf{L-k1}, where $F(X) = F_0 X^n - 2V(\phi)$, we find again
\beq                  \label{ss1}
              v_s^2 = \frac{1}{2n - 1}
\eeq
  in full agreement with \rf{v-fl}. Thus there is a complete equivalence between 
  a perfect fluid and k-essence without a potential not only for a cosmological 
  background but even on the perturbative level as far as adiabatic perturbations
  are concerned. In particular, if $w < 0$, corresponding to $n < 1/2$, the model 
  is perturbatively unstable since it implies $v_s^2 < 0$. This is true both for a perfect 
  fluid and for k-essence. Moreover, although the presence of a potential 
  changes the scalar field dynamics, the propagation speed of its perturbations, coinciding 
  with the derived speed of sound \cite{neven}, is still the same as with $V=0$.  
  
  In Rastall's theory things may be different. The speed of sound for a scalar field is 
  given by \cite{oliver, liddle}
\beq                    \label{ss2}
                 v_s^2 = \frac{2 - b}{b}.
\eeq
  In scalar vacuum and in the R1 case (matter obeys the non-conservation equation 
  \rf{cons-i1}), we have the relation \rf {nb}, $(2-b)n = 1$, which makes (\ref{ss1}) 
  and (\ref{ss2}) identical. Furthermore, the fluids in the corresponding models 
  obey different equations of state, see \rf{id-m}. However, in the Rastall model we 
  can still characterize the fluid by the ``effective'' density and pressure, $\trho$ and 
  $\tp$, the SET written in their terms is conservative, hence the squared speed of 
  sound of the Rastall fluid is equal to $d\tp/d\trho = d p_k/d\rho_k$.
  Thus we can conclude, even without performing a complete perturbation analysis, 
  that the models belonging to the two theories coincide not
  only at the background level but also at the level of adiabatic perturbations.

  In the case R2 (conserved matter), we have another relation \rf{nb-ii} between the 
  parameters $b$ and $n$, without such a simple connection. As a consequence, in principle 
  it is possible that an unstable model in a k-essence model may correspond to a stable model 
  in Rastall's theory, or vice versa, since, as shown above, \eq \rf{nb} does not hold,
  $b$ being now essentially independent of $n$ up to some possible restrictions on their range. 
  In fact, in this case, even non-adiabatic perturbations may appear, due to the coupling
  between the scalar field and matter. 

\section{Some special cases}    

\subsection{$n=1/2$}    
  
  In this case, the k-essence scalar field equation takes the form
\beq                         \label{n-half}
		3H = -2 F_0^{-1} V_\phi. 
\eeq
  Thus if $V=\const$, we have $H=0$, hence $a = \const$, and flat space-time is obtained. 
  One can certainly obtain $H=0$ in Rastall gravity under special assumptions, but the question
  of k-R duality looks meaningless in this trivial case. 
  
  If $V=V(\phi)$, the $\phi$ field has no dynamics of its own, but \eqn{n-half} expresses it
  in terms of $H$, and the Friedmann equations \rf{tt-k1} and \rf{ii-k1} are meaningful.
  
  The Rastall counterpart in the scalar-vacuum and R1 cases is then obtained with $b=0$, 
  $V = 3W$ (according to \rf{nb} and \rf{VW}) and 
\beq                                                                          \label{psif-half}
		2\ep \dot\psi{}^2 = F_0 \dot\phi. 
\eeq

  In the case R2 (conserved matter), $b$ remains arbitrary, but k-R duality still holds.
  Indeed, if we substitute \rf{psif-half} into \eqn{epsi-ii} and use the Friedmann-like 
  equations \rf{tt-Ra}, \rf{ii-Ra} to calculate $\dot\rho-3\dot p$, we obtain \eqn{n-half}.

  The main feature in this case is that nontrivial solutions with $n = 1/2$ and k-R duality
  are only achieved in the presence of a variable potential.

\subsection{$b=3/2$}

  With this value of $b$, the potential disappears from Rastall's gravity, hence the k-R duality 
  implies $V=0$, and we deal with zero potentials. In other respects, the situation is  
  described as in the general case. 

  A feature of interest is that with $b = 3/2$ \eqn{redef} leads to $\trho = 3\tp$. 
  Therefore, in the scalar-vacuum and R1 cases, the dual k-essence counterpart of 
  this Rastall model contains matter with $\rho_k= 3p_k$ (see \rf{id-m}).  
  Due to \rf{nb}, in addition, $n=2$, so that the $\phi$ field also behaves as radiation.

  In the case R2 (conserved matter), the relations \rf{id-m} and \rf{nb} are no more 
  valid, and the general description is applicable.  
      
\subsection{$b=2$}

  Equations \rf{w} and \rf{ss2} give zero values of pressure and the speed of sound of
  a scalar field in Rastall's theory. The corresponding expression \rf{ss1} in k-essence 
  theory leads to $n \to \infty$ according to the general relation \rf{nb}.
 
  If there is conserved matter (case R2), then (unless this conserved matter is pure
  radiation, $\rho = 3p$) \eqn{nb} is no more valid, so that the speeds of sound 
  of scalar fields are different in the two theories. It means that k-R duality does not
  exist for perturbations even though it does exist for the isotropic background.

\subsection{$b=0$}

  In the scalar-vacuum and R1 cases we return to the above description for $n = 1/2$.
  
  With conserved matter (case R2), the Rastall scalar field takes the form
\beq                 \label{psi-b0}
         3H \ep \dot\psi{}^2 = -3 \dot W -\Half (\dot\rho-3\dot p),
\eeq
  looking like a constraint equation since it contains only the first-order derivative.
  However, the k-R duality still works, as before: thus, a substitution of \rf{psif} 
  (what is important, with arbitrary $n \ne 0$) and $\rho-3p$ from \eqs \rf{tt-Ra}, 
  \rf{ii-Ra} into \rf{psi-b0} leads to \rf{phi-k1}, which is a second-order equation 
  unless $n = 1/2$.  

\section{Examples}

  Let us now consider some specific examples of the equivalence discussed above,
  assuming a zero or constant potential and dust as a possible matter contribution.

\subsection{Scalar vacuum}

  Consider scalar vacuum with zero potential. The k-essence equations give
\bear                                    \label{sfe}
	\dot\phi \eql  \phi_0  a^{- 3/(2n - 1)},    \qquad  \phi_0 = \const,
\yy
          H^2 \eql  \frac{2n - 1}{6}F_0 \phi_0^{2n}a^{-6n/(2n - 1)}.
\ear
  In terms of cosmic time we obtain
\bear
               a \eql a_0  t^{2/[3(1 + w)]},  \qquad  a_0 = \const,
\yy
	  \dot\phi \eql  \phi_1  t^{- 2w/(1 + w)},
\ear
  where $\phi_1$ is a combination of the previous constants, and we 
  have written $w = 1/(2n - 1)$, thus identifying the k-essence with a perfect 
  fluid with the EoS $p = w\rho$. 

  In Rastall's theory, the dual solution contains the same $a(t)$, while the scalar field is given by
\beq                              \label {sfr}
                 \dot\psi \propto a^{- 3(1 + w)/2} = a^{-3/b} \propto t^{-1},
\eeq
  where now we should put $w = (2-b)/b$. We notice that while the k-essence scalar field 
  behavior depends on the EoS parameter $w$, the Rastall scalar is simply $\psi = \log t + \const$.

  Addition of a constant potential, equivalent to a cosmological constant, does not change the 
  scalar field evolution laws \rf{sfe} and \rf{sfr} in terms of $a$ but makes the time 
  dependences more complex, not to be considered here.

  In the presence of matter, as we saw above, the form of k-R duality depends on how matter 
  couples to the scalar field. 

\subsection{Dust and Rastall-R1 models}

  Suppose that in k-essence theory, in addition to the scalar field $\phi$, there is 
  pressureless fluid (dust), so that
\beq                            \label{dust-k} 
              p_k =0, \qquad  \rho_k = \rho_1 a^{-3}, \qquad   \rho_1=\const.
\eeq
  then, in the R1 version of Rastall's theory, according to \rf{id-m}, we have the conditions
\bear                           \label{ex1}
	 \Half\Big[(3 - b)\rho + 3(b-1)p\Big] \eql \trho = \rho_k,
\nn
          \Half \Big[(b-1)\rho + (5-3b)p \Big] \eql \tp =  0,
\ear
  leading to the following relations for the density and pressure 
\bear                               \label{ex2}
          \rho \eql \frac{5-3b}{2(3-2b)} \rho_k,
\nn
	p \eql \frac{1-b}{5-3b} \rho = \frac{1-b}{2(3-2b)} \rho_k,
\ear
  both evolving as $\rho \propto p \propto a^{-3}$. Thus in Rastall cosmology the 
  fluid acquires pressure (except for the GR value $b = 1$). The scalar fields in both 
  models satisfy the same relations as in the vacuum case, valid for any values of 
  $n$ and $b$ such that $n(b-2)=1$. 

  Adding a constant potential $V = (3-2b)W$ does not change the relations \rf{ex1}
  and \rf{ex2} and introduces an effective cosmological constant. We then obtain  
  a three-component model with matter, a cosmological constant and a scalar field 
  whose behavior is determined by $n$ or, equivalently, by $w = 1/(2n-1)$. In the  
  dual Rastall model, we have an effective pressure even though in the k-essence 
  model $p_k=0$. 

  If, on the contrary, we introduce matter with $p=0$ in Rastall's (R1) theory, 
  then in the dual k-essence model we have 
\beq
              \rho_k = \trho = \Half (3-b)\rho, \qquad   p_k = \tp = \Half (b-1)\rho, 
\eeq
  and their evolution law agreeing with \eqn{cons-i} reads
\beq
              \rho_k \propto \rho \propto a^{-6/(3-b)} = a^{-3(1+w_k)},                 
\eeq
  where the EoS parameter $w_k$ of the fluid in the k-essence model is 
\beq
              w_k = \frac{b-1}{3-b} = \frac{n-1}{n+1}       
\eeq
  (not to be confused with $w = 1/(2n-1)$ characterizing the $\phi$
  field behavior); as before, the relation $n(2-b) =1$ holds. The model thus obtained
  is quite different from the one with dust introduced in k-essence theory.

\subsection{Dust and Rastall-R2 models}

  Let us again assume \eqn{dust-k} but now consider version R2 of Rastall's theory,
  so that now $\rho \propto a^{-3}$ and 
\beq                    \label{nb-ex}
	\ep \dot\psi^2\Big[b - \frac{2n - 1}{n}\Big] = (b - 1)\rho \propto a^{-3}.
\eeq
  Then for $b \neq 1$ we find according to \rf{psif-ii}
\bear
	\dot\psi &\propto& a^{- 3/2},
\nn
           \dot\phi^n &\propto& a^{-3/2} \ \ \then \ \ \dot\phi \propto a^{-3/(2n)}.
\ear
  Combining this with the relation $\ep \dot\psi{}^2 = nF_0 \dot\phi{}^{2n}$ and the 
  field equation \rf{phi-k} with $V_\phi=0$, we find that this situation corresponds to
  the limit $n \to \infty$, which is, however, well defined. In this way we obtain 
  $a \propto t^{2/3}$ as in the pure dust model of GR. 
  For the scalar fields it follows in this limit
\bearr 
        \dot\phi =\const \ \then \ \phi \propto t,
\nnn
        \dot\psi \propto a^{-3/2} \propto 1/t.
\ear
  The condition for $b$ is obtained from \rf{nb-ex}: writing $\dot\psi = \psi_0/t $ and
  $\rho = \rho_0/t^2$, we arrive at
\beq
	\ep\psi_0^2 (b - 2) = (b - 1)\rho_0.
\eeq
  Thus the value of $b$ is determined by the relative contributions of matter and the scalar field. 
  Moreover, the speed of sound of the scalar field now does not follow the adiabatic relation 
  verified in the R1 case.

  A cosmological constant can be easily introduced in the form of $V = (3-2b) W = \const$. 
  The scalar field again follows the law \rf{nb-ex}, and the whole configuration reduces to the
  $\Lambda$CDM model where $\Lambda$ is given by the constant potential and the matter 
  component consists of the scalar field and ordinary matter. All background relations of the 
  $\Lambda$CDM model are preserved in this case, but the degeneracy between the scalar 
  field and usual matter is broken at the perturbative level. Due to the fact that the 
  $\Lambda$CDM model is subject to problem at the perturbative level in the non-linear regime 
  (see, e.g., \cite{cusp,satt}), such a more complex configuration in k-essence and Rastall 
  models may lead to interesting results, to be studied in the future.

\section{Conclusion}

  We have studied the conditions of equivalence between the k-essence and Rastall theories of 
  gravity in the presence a scalar field (k-R duality). These two theories have actually emerged in 
  very different contexts, the k-essence theory being based on a generalization of the kinetic term 
  of a scalar field, suggested by some fundamental theories, while Rastall's theory is a 
  non-conservative theory of gravity which can be seen as a possible phenomenological 
  implementation of quantum effects in gravitational theories. Such equivalence has been 
  revealed in the case of static spherically symmetric models \cite{denis, we-16}, and it 
  has been more explicitly stated here for all cases where the metric and scalar fields essentially
  depend on a single coordinate, and the k-essence theory is specified by a power-law function 
  of the usual kinetic term, to which a potential term can be added. This generalization covers 
  diverse static and cosmological models, including all homogeneous cosmologies.

  We have discussed cosmological configurations with scalar fields and matter in the form of 
  a perfect fluid whose evolution in Rastall's theory can follow one of two possible laws: one (R1) 
  assumes no mixing between matter and the scalar field, each of them separately obeying 
  the non-conservation law \rf{cons-R}, and the other (R2) ascribes the whole non-conservation 
  to the scalar field while matter is conservative ($\nabla_\nu T\mN[m] =0$). Let us summarize
  the main results obtained in this context:
\begin{enumerate}
\item 
	k-R duality has been established for version R1 of Rastall's theory with an arbitrary 
 	EoS of matter. It has been found that the EoS of matter is different in the mutually dual
  	k-essence and Rastall models; however, it is argued that the respective speeds of 
	sound are the same. Since the speeds of sound characterizing the scalar fields ($\phi$ in 
	k-essence theory and $\psi$ in Rastall's) also coincide, we conclude that k-R duality
         is maintained not only for the cosmological backgrounds but also for adiabatic 
         perturbations.
\item  
	For version R2 of Rastall's theory, it has been found that k-R duality exists only with 
	fluids having the EoS $p = w\rho,\ w=\const$, which is the same for k-essence and 
	Rastall models. Moreover, in the k-essence model the scalar field obeys the same 
	effective EoS. However, on the perturbative level the mutually dual models 
	behave, in general, differently. 
\item 
	Some special cases have been discussed, showing how there emerge some restrictions 
	on the free parameters of each theory. 
\item 
	An example has been considered in which a cosmological model completely equivalent 
         to the $\Lambda$CDM model of GR is obtained at the background level, but different 
	features must appear at the perturbative level. 
\end{enumerate}

  The equivalence between the two theories discussed here is somewhat surprising because of
  their basically different origin. A curious aspect is that the k-essence theory has a Lagrangian
  formulation unlike Rastall's theory. It is possible that the equivalence studied here may lead to
  a restricted Lagrangian formulation of Rastall's theory in the minisuperspace in terms of 
  metric functions depending on a single variable. If this is true, it might suggest how to recover 
  a complete Lagrangian formulation for Rastall's theory in a more general framework.

\subsection*{Acknowledgments} 

  We thank CNPq (Brazil) and FAPES (Brazil) for partial financial support.
  KB thanks his colleagues from UFES for kind hospitality and collaboration.
  The work of KB was performed within the framework of the Center 
  FRPP supported by MEPhI Academic Excellence Project 
  (contract No. 02.a03. 21.0005, 27.08.2013),
  within the RUDN University program 5-100, and RFBR project No. 16-02-00602.

\end{document}